\documentclass[12pt,preprint]{aastex}

\slugcomment{Accepted to ApJ (July 1, 2013)}

\shorttitle{L1521F-IRS}
\shortauthors{Takahashi et al.}

\begin{document}

\title{Direct Imaging of a Compact Molecular Outflow from a Very
Low-luminosity Object; L1521F-IRS}

\author{Satoko Takahashi}
\affil{Joint ALMA Observatory, Alonso de Cordova 3108, Vitacura, Santiago, Chile; \\
National Astronomical Observatory of Japan, 2-21-1 Osawa, Mitaka, Tokyo 181-8588, Japan; \\
Academia Sinica Institute of Astronomy and Astrophysics, P.O. Box
23-141, Taipei 10617, Taiwan \\
$satoko.takahashi@nao.ac.jp$\\
}

\author{Nagayoshi Ohashi}
\affil{Subaru Telescope, National Astronomical Observatory of Japan, North
Aohoku Place Hilo HI 96720, USA}
\affil{Academia Sinica Institute of Astronomy and Astrophysics, P.O. Box
23-141, Taipei 10617, Taiwan}

\author{Tyler L. Bourke}
\affil{Harvard-Smithsonian Center for Astrophysics, 60 Garden Street
Cambridge, MA 02138, U. S. A.}

\begin{abstract}

Studying the physical conditions of very low-luminosity objects (VeLLOs;
$L_{\rm{bol}}<0.1$ L$_{\odot}$) is important for understanding the
earliest evolutionary stage of protostars and brown dwarfs.  We report
interferometric observations of the VeLLO L1521F-IRS, in \hbox{$^{12}$CO
(2--1)} line emission and the \hbox{1.3 mm} continuum emission, using the
Submillimeter Array \hbox{(SMA)}.  With the \hbox{$^{12}$CO (2--1)}
high-resolution observations, we have spatially resolved a compact but
poorly collimated molecular outflow associated with \hbox{L1521F-IRS} for
the first time.  The blueshifted and redshifted lobes are aligned along the
east and west side of \hbox{L1521F-IRS} with a lobe size of
\hbox{$\approx$1000 AU}.  The estimated outflow mass, maximum outflow
velocity, and outflow force are \hbox{(9.0-80)${\times}$10$^{-4}$ M$_{\odot}$},
7.2 km s$^{-1}$, and \hbox{(7.4-66)$\times$10$^{-7}$ M$_{\odot}$ km s$^{-1}$
yr$^{-1}$}, respectively.  The estimated outflow parameters such as size,
mass, and momentum rate are similar to values derived for other VeLLOs, and
are located at the lower end of values compared to previously studied
outflows associated with low- to high-mass star forming regions.
Low-velocity less collimated (1.5 km s$^{-1}$/1200 AU) and higher-velocity
compact (4.0 km s$^{-1}$/920 AU) outflow components are suggested by the
data.  These velocity structures are not consistent with those expected in
the jet driven or wind driven outflow models, perhaps suggesting a
remnant outflow from the FHSC as well as an undeveloped
outflow from the protostar.  Detection of an infrared source and compact
millimeter continuum emission suggest the presence of the protostar, while
its low bolometric luminosity (0.034--0.07 L$_{\odot}$), and small outflow, 
suggests that L1521F is in the earliest protostellar stage ($<10^{4}$ yr) and
contains a substellar mass object. 
The bolometric (or internal) luminosity of L1521F-IRS suggests that the current 
mass accretion rate is an order-of-magnitude lower than expected in the standard
mass accretion model ($\approx$10$^{-6}$ M$_{\odot}$
yr$^{-1}$), which may imply that L1521F-IRS is currently in a low activity
phase.  

\end{abstract}

\keywords{ISM: clouds --- ISM: individual objects (L1521F-IRS, Taurus) ---stars: jet and outflow ---
stars: protostars ---ISM: molecules---stars: formation---stars: low-mass}

\section{INTRODUCTION}

	Studying the earliest phase of star formation is an important issue
for understanding the initial conditions of star formation.  Previous
studies of star formation have focused on dense cores associated with
\hbox{class I/0} protostars, whereas these class I/0 objects are rather
well-developed protostars with estimated ages of \hbox{10$^4$--10$^5$ yr}.
According to the conventional accretion scenario of protostars (e.g.,
Stahler et al. 1980), in which the accretion luminosity is given as
\hbox{$GM_{\ast}{\dot{M}_{\rm{acc}}}/R_{\ast}$} ($G$ is the gravitational constant,
$\dot{M}_{\rm{acc}}$ is the mass accretion rate, and $M_{\ast}$ and $R_{\ast}$ are the
mass and the radius of the protostar, respectively), extremely young
protostars in a very early phase of the accretion process are less luminous
when compared with well-developed protostars. Such extremely young
protostars, though their total number would be small because of their short
time scale, have most probably not been identified in many previous
studies. In order for us to study the initial conditions, or extremely early
phase of star formation, it is crucial for us to pay more attention to
starless cores, where no bright infrared sources are found.  Starless cores
with high central densities \hbox{(${\geq}10^{5}$ cm$^{-3}$)} are sometimes
called pre-stellar cores, as they show signs of evolution toward the
formation of a protostar, such as inward motions (e.g., \hbox{L1544};
Tafalla et al 1998. Ohashi et al. 1999). 

	Great progress in the study of starless cores has been made using
the {\it Spitzer} Space Telescope, which archived unprecedented sensitivity
($L<1$ L$_{\odot}$) in nearby star-forming regions.  Spitzer observations
have identified a large number of low-luminosity candidates, including 15
Very Low Luminosity Objects, or VeLLOs ($L_{\rm bol} < 0.1 L_{\Sun}$; Di
Francesco et al.\ 2007, Bourke et al. 2005, 2006; Dunham et al. 2008;
Terebey et al. 2009).  The first VeLLO, \hbox{L1014-IRS}, was discovered at
the center of a ``starless" core with a bolometric luminosity of 0.09
L$_{\odot}$ (Young et al. 2004).  Follow up SMA observations toward VeLLOs, L1014-IRS and L1451-mm, 
have detected outflows associated with them. The estimated outflow parameters suggest
that VeLLOs have the most compact, the lowest mass and
the least energetic outflows compared with known Class 0/I outflows (Bourke et al.
2006; Pineda et al. 2011).  These results suggest that L1014-IRS would be a
very low-mass protostar (i.e., proto-brown dwarf star) and L1451-mm is a
candidate of First Hydro-Static Core (FHSC).  Recent detailed case studies of VeLLOs
or similar targets present some variety in their nature such
as the mass of parental core, outflow parameters, and outflow morphology
(Bourke et al. 2005; 2006; Terebey et al. 2009; Belloche et al. 2002; 2006;
Chen et al. 2010; Dunham et al. 2010; Pineda et al. 2011; Kauffmann 2011;
Dunham et al. 2011).  However, the nature of VeLLOs are not clear yet due
to their limited in number. 

There are three possible scenarios for the nature
of VeLLOs.  Some source are proto-brown dwarfes, extremely
young protostars, or young stellar objects, which are in a low-state of accretion phase 
(not necessarily ``extremely young''). 
Studying these candidates is very important because for
sources related to the first type are crucial to investigate the formation
and evolution of brown dwarfs. Sources of the second type are related to
the earliest stage of star formation, and provide a missing link between 
low-mass star formation at a very earlier evolutionary phase, finding a
missing link between the prestellar phase and the Class 0 phase.  Theories
predict that the earliest protostellar phase is the important FHSC phase, 
followed by the subsequent formation of a protostar (Larson
1969; Masunaga \& Inutsuka 2000; Bate 1998, 2010; Saigo et al.  2008;
Schonke \& Tscharnuter 2011).  Studies of VeLLOs are crucial to
observationally test the protostar formation scenario through FHSC. 
Sources of the third type are crucial to understand how the accretion process terminate 
at the last phase of the young stellar evolution or how the accretion can be periodic. 

	In order to study nature of VeLLOs, in this paper we focus on a 
detailed case study of the VeLLO, \hbox{L1521F-IRS}, which is located in
the Taurus molecular cloud ($d=$140 pc; $L_{\rm{IR}}=$0.024 L$_{\odot}$).
The parental core, \hbox{L1521F} (or \hbox{MC 27} by Mizuno et al. 1994), was originally
noticed as a starless condensation with a high-central density of
$\approx$10$^6$ cm$^{-3}$ (Onishi et al. 1998), and lately identified as
the second example of a VeLLO (Bourke et al. 2006). 
Molecular depletion observed in CCS and the enhanced deuterium fraction imply that 
L1521F is at the time one of the two best examples of an evolved starless core in the Taurus
molecular cloud ($d=140$ pc), along with \hbox{L1544} (Crapsi et al. 2004;
2005; Shinnaga et al. 2004). In contrast to another VeLLO core, \hbox{L1014}, \hbox{L1521F} was
identified as the most evolved starless core, may be suggesting that they are 
extremely young protostars. 
Asymmetric spectra observed in HCO$^{+}$ (4--3 and 3--2) emission, suggest
infalling gas motions with a size scale of \hbox{2000-3000 AU} (Onishi et al. 1999). 
The {\it Spitzer}
observations discovered a reflection nebula and a point source associated
with L1521F (Bourke et al. 2006). 
The estimated luminosity is 0.034--0.07 L$_{\odot}$ (Bourke et al. 2006; Terebey et al. 2009). 
Detection of a 100 AU scale dust continuum source with 1.3 mm PdBI observations 
(Maury et al. 2010) supports the claim that the protostar
has already formed at the center of L1521F.  Recent single-dish studies in
the CO (7--6 and 6--5) emission detected warm ($\sim$30-70 K) and extended
($\sim$2400 AU) gas, suggesting that these emission may be originated from 
shock gas at the interface between the outflow and dense core (Shinnaga et al. 2009). 	

	In this paper, we present the \hbox{1.3 mm} dust continuum and
\hbox{$^{12}$CO (2--1)} results obtained from the Submillimeter Array (SMA;
Ho et al. 2004). Goals of the paper are to (i) search for a compact
molecular outflow associated with \hbox{L1521F-IRS}, (ii) investigate the
physical properties of the molecular outflow, (iii) compare its properties
with molecular outflows detected from other candidate VeLLOs, and (iv)
discuss the evolutionary stage of L1521F-IRS. 
		
	The SMA observations and data reduction are explained in
\hbox{Section 2}. The synthesized SMA images, derived physical parameters,
spectra and kinematics of \hbox{CO (2--1)} emission and the \hbox{1.3 mm}
continuum emission are presented in \hbox{Section 3}.  Outflow properties,
evolutionary phase, and the origin of L1521F-IRS are discussed in Section 4.
Finally \hbox{Section 5} summarizes the project.

\section{OBSERVATIONS AND DATA REDUCTION}

		The \hbox{$^{12}$CO (2--1; 230.538 GHz)}, $^{13}$CO (2--1;
220.39868 GHz), C$^{18}$O (2--1; 219.56036 GHz), N$_2$D$^{+}$ (3--2;
231.321 GHz) and \hbox{1.3 mm} continuum observations were carried out
with the eight 6 m antennas of the SMA\footnote{The
Submillimeter Array is a joint project between the Smithsonian
Astrophysical Observatory and the Academia Sinica Institute of Astronomy
and Astrophysics and is funded by the Smithsonian Institution and the
Academia Snica.} in its compact configuration on January 3, 2007. 
The phase center was set at R.A.(J2000) = 4$^h$28$^m$38.95$^s$,
decl.(J2000)=26$^{\circ}$51$'$35$''$.1. 
Zenith opacities at \hbox{225 GHz} were typically in the range between
\hbox{0.18--0.2}.  The typical system noise temperature in DSB mode was
between 100--200 K.  Both the LSB and USB data were
obtained simultaneously by the digital spectral correlator, which has a bandwidth of
\hbox{2 GHz} in each sideband. 
The SMA correlator was configured with high spectral
resolution windows of 256 channels over \hbox{104 MHz} for \hbox{$^{12}$CO
(2--1)} and $^{13}$CO (2--1), and 1024 channels over 104 MHz for C$^{18}$O
(2--1) and N$_2$D$^{+}$ (3--2).  These provide channel spacing of
\hbox{0.53 km s$^{-1}$} and \hbox{0.13 km s$^{-1}$}, respectively. 
After subtracting the spectral windows where CO
(2--1), $^{13}$CO (2--1), and C$^{18}$O (2--1) are located at, the LSB and USB
continuum data, separated by 10 GHz, were combined to improve
the sensitivity. The effective bandwidth for the continuum emission is
approximately 3.0 GHz. 
		
		The phase and amplitude were calibrated by observations of 3C 111 \hbox{(3.0
Jy)}, while the passband calibration was made through observations of the
quasar 3C 279.  The flux calibration was performed by observations of
Uranus.  The overall flux uncertainty was estimated to be
\hbox{$\sim$20\%}. 
The array provided projected baselines ranging from \hbox{9 to 53 k$\lambda$},
so that our observations were insensitive to structure larger than
\hbox{$18''$} (equivalent to \hbox{2500 AU}) at the \hbox{10 \%} level
(Wilner \& Welch 1994),  but have good sensitivity to spatial scales
below this.  The primary FWHM beam of the SMA is \hbox{${\sim}55''$} at
this frequency. 
				
		The raw data were calibrated using MIR, originally
developed for the Owens Valley Radio Observatory (OVRO; Scoville et al.
1993) and adopted for the SMA. After the calibration, final CLEANed images
were made with the AIPS task IMAGER. 
In order to image the extended faint continuum emission, Natural weighting  with a Gaussian taper (FWHM=30 k$\lambda$, corresponding to 6$''$) 
was used for the 1.3 mm continuum data, while natural weight with no taper was used for the strongly detected $^{12}$CO (2-1) data. 
The resultant angular resolutions of the
images are $4''.8{\times}4''.4$ and $3''.1{\times}2''.7$ for the 1.3 mm
continuum and \hbox{$^{12}$CO (2--1)} data, respectively.  The achieved rms
noise levels of the \hbox{1.3 mm} continuum and $^{12}$CO (2--1) images are
\hbox{1.3 mJy beam$^{-1}$}, and \hbox{0.13 Jy beam$^{-1}$} in a \hbox{0.53
km s$^{-1}$} channel, respectively.  Note that the $^{13}$CO (2--1) line
was barely detected (${\leq}5{\sigma}$), so the data were only used for
the opacity estimation described in Section 3.3.2.  No C$^{18}$O (2--1) nor
N$_2$D${+}$ (3--2) emission was detected even after smoothing the channels.
The observational parameters for CO (2--1) and 1.3 mm continuum are
summarized in \hbox{Table 1}.

\section{RESULTS} 			
\subsection{1.3 mm Continuum Emission}

	Figure 1 shows the \hbox{1.3 mm} continuum emission toward 
\hbox{L1521-IRS} detected with the SMA (white contours), superposed on the
\hbox{1.2 mm} continuum emission obtained with the IRAM 30 m/MAMBO-2 
(black contours and grey scale).  
We have detected the \hbox{1.3 mm} continuum emission
with a peak position of R.A.(J2000) = 4$^h$28$^m$38.82$^s$,
decl.(J2000)=26$^{\circ}$51$'$34$''$.1. 
The total flux density and peak intensity
are measured to be \hbox{11 mJy} and \hbox{6.8$\pm$1.3 mJy
beam$^{-1}$}, respectively. 
The position and peak flux were derived with the AIPS task ``imstat'', while the total flux was 
derived from two dimensional Gaussian fitting with the AIPS task ``imfit''. 
The peak position of \hbox{1.3 mm} continuum 
is roughly consistent with previous single-dish results in the
(sub)millimeter continuum emission (Shinnaga et al 2004; Crapsi et al.
2004; Kauffmann et al. 2008). 
The structure of the 1.3 mm continuum
emission is elongated in the east-west direction with a deconvolved size of
$5''.0$ (700 AU), while the structure along the north-south direction is
not spatially resolved with our SMA observations.  

	On the assumption that the dust emission is optically thin at
\hbox{1.3 mm} and the temperature distribution of the dust-continuum
emission is uniform, the mass of the dust condensation is estimated to be
\hbox{0.047 --0.084 M$_{\odot}$}, and the lower limit of molecular hydrogen
column density is estimated to be \hbox{(1.3--2.3)$\times$10$^{23}$
cm$^{-2}$}\footnote{The upper size of the continuum source, which is
\hbox{700$\times$620 AU}, was derived from the deconvolved size of the
major axis and the beam size of the minor axis using the 2D Gaussian
fitting tool AIPS ``imfit'', and then, the mean number density was
estimated with the assumption of an elliptical shaped structure}.  For the
derivation, we assumed that \hbox{$\beta$=1.5--2.0}, a dust temperature of
\hbox{10 K} (e.g., Myers \& Benson 1983), a gas-to-dust ratio of 100, and $\kappa_0$ = 0.0899
cm$^{2}$ g$^{-2}$ at 1.3 mm (Ossenkopf \& Henning 1994). 
When a dust temperature of 30 K instead of 10 K is adopted, the estimated mass 
becomes larger by a factor of 5.
The flux density of the 1.3 mm continuum emission measured with the SMA corresponds to
\hbox{7.6 \%} of the flux density measured within the \hbox{10$''$.5} beam
of the \hbox{IRAM/MAMBO-2} \hbox{1.2 mm} observations, which is \hbox{90
mJy} (Crapsi et al. 2004). 

	Although the SMA 1.3 mm continuum emission is associated with the
Spitzer source, its peak position is shifted by \hbox{2.$''$0} from the mean
position of the Spitzer continuum bands in which infrared source is
detected (wavelength longer than the 5.8 $\mu$m band; Bourke et al. 2006).
Since near infrared emission in the {\it Spitzer}/IRAC image may be
attributed mainly to scattered light, we have also checked the position of
the \hbox{24 $\mu$m} source taken with the {\it Spitzer}/MIPS; it is found
that the peak position of \hbox{24 $\mu$m} emission is much closer to the
mean position of {\it Spitzer}/IRAC image than the SMA \hbox{1.3 mm}
continuum source position.  Since the positional shift between the
\hbox{1.3 mm} and infrared sources is approximately 9 times larger than
the positional accuracy of the SMA observations, 0\farcs23, 
the positional shift between the \hbox{1.3 mm} continuum source and the
infrared source is significant. 
Here, the positional accuracy of the SMA observations was derived from combination of the signal-to-noise
ratio (S/N) of source and the beam size relation: ${\sigma_{\rm{SN}}}{\approx}(1/2{\pi})({\theta}/(S/N)){\approx}0''.19$, and baseline error of 
the SMA observations: ${\sigma_{\rm{BL}}}{\approx}0.1{\lambda}{\approx}0''.13$; which is ${\approx}{\sqrt{{{\sigma_{\rm{SN}}}^2}+{{\sigma_{\rm{BL}}}^2}}}{\approx}0\farcs23$.

	Observations of L1521F using the Plateau de Bure
Interferometer (PdBI) at a angular resolution of $0''.49{\times}0''.27$ 
by Maury et al. (2010) in 1.3 mm continuum emission detected a compact source 
associated with the Spitzer source. The
reported peak flux of 1 mJy beam$^{-1}$ (8.3$\sigma$ detection) by PdBI is much
weaker than the peak flux of the continuum emission detected by SMA, and
the deconvolved size of 0.65$''$ ($\sim$90 AU) is much more compact than
the SMA continuum emission. The origins of these continuum sources detected
with the SMA and PdBI are discussed in Section 4.1.

\subsection{First Detection of Compact Molecular Outflow associated with
\hbox{L1521F-IRS}}

	The $^{12}$CO (2--1) emission was detected at blueshifted
(V$_{\rm{LSR}}$= \hbox{2.9--6.1 km s$^{-1}$)} and redshifted
\hbox{(V$_{\rm{LSR}}$=8.2--10.3 km s$^{-1}$)} velocities in our SMA
observations.  Figure 2 shows integrated intensity maps of the blueshifted
and redshifted emission, superposed on the 4.5 $\mu$m image obtained with 
the \hbox{{\it Spitzer}/IRAC}. This is the first time that the CO outflow emission associated with L1521F-IRS has
been spatially resolved and their distributions are directly
compared with the reflection nebula obtained with \hbox{{\it
Spitzer}/IRAC}.  The CO emission extends approximately $18''$ (\hbox{2500
AU}) with the signal-to-noise level of more than three.  The near infrared reflection nebula as well as the high-extinction
region around L1521F-IRS shown in Figure 2a suggest that the
axis of the outflow cavity is aligned at a position angle of $75^{\circ}$
(denoted as dashed line in Figure 2a). The detected CO emission is
elongated almost in the east-west direction, which is similar
to the near-infrared nebula. 
Careful inspection reveals that the infrared nebula is slightly brighter to the west of 
\hbox{L1521F-IRS}, suggesting that the western part of the reflection nebula is
slightly tilted to the near-side from the plane of the sky (i.e., envelope inclination angle of 50$^{\circ}$--70$^{\circ}$; Terebey et al. 2009).  
Furthermore, the reflection nebula indicates a wide opening angle cavity, hence it is
expected that both blue and redshifted CO emission appears in both outflow
lobes. The observed CO (2--1) emission shows more complicated structures than
the reflection nebula.  The blueshifted CO emission is brighter on the
eastern side, which is in the opposite sense  to what might be expected
from the reflection nebula image. This may be a consequence of the
non-uniform distribution of the surrounding material. The absence of the
redshifted emission can be explained by the absorption from foreground
diffuse gas, as mentioned in Section 3.3.  The CO integrated emission shows
a parabolic shaped structure, especially in the blueshifted velocity 
detected in the eastern part of the emission (Figure 2a and 2b). 

	Further, as presented in Figure 2c, compact CO blue and redshifted
components located to the north-east of L1521F-IRS are distributed symmetrically 
with respect to a point denoted by an open star.  This might suggest that the SMA
CO emission traces multiple outflows from a binary system (one associated
with L1521F-IRS and another one associated with a new source, L1521F-NE).
However, no driving source has been detected in either millimeter continuum emission 
with PdBI/SMA or infrared emission with Spitzer the open star position (L
1521F-NE), with a mass detection limit of 0.0005 M$_{\odot}$
\footnote{The mass detection limit are derived from the 3$\times$rms noise
level obtained with the PdBI observations by Maury et al. (2010) with
assumptions of $\beta$=1.5 and T=20K (refer more detail in Section 3.1)}
(Maury et al. 2010; This work).  
Since there is no strong evidence to support the 2nd case, 
in this paper, we consider that a single outflow is driven by L1521F-IRS, and
discuss the velocity structures and physical parameters as the single outflow. 

	Figure 3 shows velocity-channel maps of \hbox{$^{12}$CO (2--1)}
emission, with a velocity interval of \hbox{0.53 km s$^{-1}$}.  Here, we
adopt the systemic velocity of \hbox{6.5 km s$^{-1}$} obtained from the
single-dish \hbox{N$_2$D$^+$} spectrum measured by Crapsi et al.  (2004).
In the velocity-channel maps no blueshifted or redshifted components with
velocity more than \hbox{$|V_{\rm{LSR}}-V_{\rm{sys}}|>$ 4.5 km s$^{-1}$}
were detected at a 3$\sigma$ level.  The blueshifted component in the
velocity range of \hbox{V$_{\rm{LSR}}$=4.0--4.5 km s$^{-1}$} is located at ~2" west of 
the protostar position (i.e., position of {\it Spitzer} point source as
well as the PdBI 1.3 mm source). In the velocity
range of \hbox{V$_{\rm{LSR}}$=5.1--6.1 km s$^{-1}$}, three emission peaks
surrounding the protostar are detected. No significant emission
($>3{\sigma}$) was detected close to the systemic velocity of
\hbox{V$_{\rm{LSR}}$=6.6--7.7 km s$^{-1}$}.  
This is most likely due to the lack of sensitivity to large-scale structures, mainly originated from extended emission such as an ambient envelope, 
although, as will be discussed in Section 3.3, missing flux originated from the outflow 
is less likely significant. In addition, self-absorption of the CO emission would also responsible 
for the less significant emission at $V_{\rm{LSR}}$=6.6--7.7 km s$^{-1}$ (see more detail in Section 3.3). 
The redshifted components in the velocity range of
\hbox{V$_{\rm{LSR}}$=8.2--8.8 km s$^{-1}$} are detected to the north-east
and west of the protostar position. The most redshifted components
in the velocity range of the \hbox{9.3--10.3 km s$^{-1}$} are detected 
\hbox{9$''$} west of the protostar position. 

	The CO emission is elongated in an east-west direction, which is
consistent with the extension of the near-infrared nebula observed in the
\hbox{{\it Spitzer}/IRAC} image. 
Hence, it is likely that the CO emission trace the outflow. 
In fact, the observed mean CO velocity, $\approx$2 km s$^{-1}$, is significantly 
larger than the velocity required to escape from the typical stellar mass in Taurus 
(0.8 M$_{\odot}$; Luhman et al. 2009 at 1000 AU, which is $\approx$0.7 km s$^{-1}$).  
Moreover, if the CO emission is associated with the near-infrared nebula, the intrinsic
CO velocity should be larger than the observed CO velocity. Considering with 
the inclination angle of between 20$^{\circ}$ and
40$^{\circ}$ (Terebey et al. 2009; pole-on is defined as an outflow
inclination angle of 0$^{\circ}$), velocities larger by factors of 2.9 and
1.6 would be expected.  This argument suggests that the CO emission is most
likely gravitationally unbound. 
Recent PdBI observations by Maury et al. (2010) did not detect the high
velocity CO (2--1) components.  This suggests that the high velocity CO is
relatively extended, but not compact enough to be detected with the PdBI
observations weighted toward the long baseline lengths.

\subsection{CO Spectra} 


	In order to discuss the outflow properties, we focus on the compact
CO outflow components detected with the SMA.  The nature of any extended
outflow component is unclear and requires combining the SMA data with
single-dish data to recover extended missing flux.  A full analysis of
these data are beyond the scope of this paper, and we note that no
high-velocity and extended outflow emission has been reported previously.
In this paper, we investigate the missing flux using SMA \hbox{CO
(2--1)} spectra compared with recently observed single-dish \hbox{CO
(2--1)} spectra using the Submillimeter Telescope (SMT; Ohashi et al.\ in
prep.).  

	Figure 4 presents a comparison of the SMA and SMT CO spectra at the phase center.  
For the spectral comparison, the SMA CO data have been convolved with a SMT
\hbox{32$''$} beam, and both the SMA and SMT spectra are compared in units
of brightness temperature.  \hbox{Figure 4a} compares the \hbox{CO (2--1)}
spectra between the SMT and SMA.  In order to compare the high-velocity CO
component originated from the outflow, a ``residual'' spectrum was derived
from the difference between the SMT spectrum at the protostar position and
a mean spectrum of the spectra at 30$''$ offsets from the central position
(\hbox{Figure 4b} thin line).  Here, we assume that the CO (2--1) emission
originating from the molecular outflow is dominant within
$R<$2000 AU ($<14''$).  This assumption is probably reasonable because no
large-scale molecular outflow has been reported from any previous
single-dish studies.  Further, the detection of higher transition CO (7--6
and 6--5) emission, which may originate in the shocked gas between the
outflow and dense envelope gas, also shows a similar size (Shinnaga et al.
2009). With this reasoning, the emission at positions offset from the
protostar mostly comes from emission in the protostellar envelope.
Subtraction of the mean spectrum of the offset positions from the central
spectrum is presented in Figure 4b as thin line.  The residual spectrum
shows a high-velocity component likely originating from the molecular
outflow associated with L1521F-IRS.  The residual CO emission appears in
the LSR velocity range of \hbox{4.4 to 6.5 km s$^{-1}$} and \hbox{8.0 to
9.4 km s$^{-1}$}.  Comparison between the SMA spectrum and the SMT residual
spectrum is presented in Figure 4b.  Both spectra show a similar spectral
shape with roughly the same velocity ranges.  This simple analysis suggests
that missing flux in the high-velocity compact component is not significant
in the SMA data. 

	In the SMT \hbox{CO (2--1)} spectrum, two intensity peaks were
detected with the LSR velocity at \hbox{5.9 km s$^{-1}$} and \hbox{7.5 km
s$^{-1}$}.  These velocity peaks are shifted from the cloud systemic
velocities, \hbox{$V_{\rm{sys}}=$6.5 km s$^{-1}$}. 
The CO spectrum taken with the SMT is poorly fitted
by a single-Gaussian component centered at the systemic velocity of \hbox{L
1521F} cloud, \hbox{$v_{\rm{sys}}=$6.5 km s$^{-1}$}.  One possibility to
explain this is the presence of a secondary component located along the same line of sight
with a different systemic velocities.  The line profile with a peak velocity
of \hbox{7.5 km s$^{-1}$} is only detected in the lower transitions of CO
(1--0 and 2--1; Our SMT data presented in Figure 4a) and $^{13}$CO (1--0) 
(Takakuwa et al. 2011; Ohashi et al. in prep.), which have low-critical densities of
\hbox{$n_c{\sim}10^2$ cm$^{-3}$}, but not in other dense
gas tracers such as H$^{13}$CO$^{+}$, N$_2$H$^{+}$, and N$_2$D$^{+}$
(Onishi et al. 1999; Crapsi et al. 2006; Shinnaga et al. 2004).  If the
secondary component (diffuse cloud) is located in the foreground of the \hbox{L1521F}
main cloud, and if the temperature of the second cloud is colder than
the \hbox{L1521F} main cloud, it is possible that CO emission from the \hbox{L
1521F} system is absorbed by the CO emission of the second cloud.  Note
that this idea also explains the missing redshifted emission seen in the
SMA \hbox{CO (2--1)} spectrum with a velocity range of 6.5 to \hbox{8.2 km
s$^{-1}$} (\hbox{Figure 4} thick line), as well as the sharp decrease in
emission in the redshifted velocity component observed in the
position-velocity (P-V) diagram presented in \hbox{Figure 5} (will be
discussed in \hbox{Section 3.4}).  This sharp emission decrease is likely
due to the self-absorption rather than missing flux in the SMA
observations.  If the lack of emission is due to the missing short
baselines, we should see the missing flux symmetrically centered at the
systematic velocity \hbox{(6.5 km s$^{-1}$)}.

\subsection{Velocity Structure}

		Previous outflow studies favor two different models for the
origin of outflows (e.g., Lee et al. 2000).  One is {\bf the wind-driven
model}: molecular outflows consist of swept-up gas entrained by a
wide-angle outflow.  The other is {\bf the jet-driven model}: molecular
outflows consist of ambient gas swept up by the bow shock at the jet head.
These two models show different spatial distributions as well as 
different kinematic features in a position-velocity (P-V) diagram cut along
the outflow axis. The wind-driven model shows a parabolic structure in the
P-V diagram that originates from the central star, while the jet-driven
model produces a large-velocity dispersion ($>$ several 10 km s$^{-1}$) at
the jet head (e.g., Lee et al.  2000). 

		A position-velocity (P-V) diagram of the \hbox{$^{12}$CO
(2--1)} emission cut along the axis of the molecular outflow
(\hbox{P.A.=75$^{\circ}$} passing through the {\it Spitzer}/IRAC position) is
shown in Figure 5.  Here, we compare the observed CO outflow with simple
analytical outflow models discussed in Lee et al. (2000). In the
cylindrical coordinate system, the structure and velocity of a parabolic
shaped shell can be descried as follows; 

		\begin{equation}
			z=CR^2, ~~~~~v_R=v_0R, ~~~~~v_z=v_0Z
		\end{equation}

		where $z$ is the distance along the outflow axis, $R$ is
the radial size of the outflow perpendicular to $z$, $C$ and $v_0$ are free
parameters that describe the spatial and velocity distributions of the
outflow shell, respectively.

		Note that redshifted emission with the LSR velocity range
between 6.5 to 8.0 km s$^{-1}$ was missed due to the foreground absorption,
so that in the velocity analysis, we mainly focus on the blueshifted
emission.  The outflow shell was delineated following the blueshifted
component of the CO intensity map obtained with the SMA and the {\it
Spitzer} 4.5 $\mu$m image in order to determine the free parameter $C$ (Figure
2a and 2b).  Adopting the outflow inclination angle of 30$^{\circ}$ (the
median value estimated from Terebey et al. 2009), the curvatures of the
parabolic shell surface, $C$, was derived as 0.022 arcsec$^{-1}$.  
With the values of $C$ and $i$ fixed, the model P-V diagram was compared 
with the observational results by varying the free parameter $v_0$. 
As an example, the solid line in Figure 5 is produced
by \hbox{$i=30^{\circ}$}, \hbox{$C$=0.022 arcsec$^{-1}$}, \hbox{$v_0=0.05$
km s$^{-1}$}. 
Varying $v_{0}$ can change the curvature of the PV model, presented in the solid line in Figure 5. 
If we assume that the velocity structure follows these 
simple models, a wind-driven outflow model predicts the Hubble law velocity
structure, that is the velocity increases as the emission moves away from
the central stars.  The observed outflow velocity structure toward L1521F
does not agree with the Hubble-law type velocity structure, which delineate parabolic shape in PV diagram, 
as denoted in Figure 5.  The L1521F outflow also does not show the large-velocity
dispersion at the jet head, which is expected in the jet-driven outflow. 
These results suggest that the velocity structure of the L1521F outflow
does not follow those observed in the low-mass protostellar outflow/jets in the 
Class 0 phase (e.g., Lee et al. 2000, 2007a, b, Hirano et al. 2010).

		In addition to the low-velocity extended components, a
higher velocity component has also been detected around the position of
\hbox{L1521F-IRS}.  The blueshifted emission of this component extends up
to the velocity of $|V_{\rm{LSR}}-V_{\rm{sys}}|$ = 3.6 km s$^{-1}$.  The
velocity channel maps (Figure 3) also show the high velocity blueshifted
component concentrated at the L1521F-IRS with the LSR velocity of 2.9 to
6.1 km s$^{-1}$.  Assuming that the inclination angle of the outflow is
$\sim$30$^{\circ}$, the intrinsic maximum velocity can be \hbox{${\sim}$7.2 km
s$^{-1}$}.  As we discussed in Section 3.2, this emission is
gravitationally unbound (even assuming the most massive stellar mass in
Taurus), and hence, unlikely tracing infalling or rotational motion of the
envelope gas.
 
		A gravitationally unbound compact CO component has been
observed toward a Class 0 source, B 335, with the SMA $^{12}$CO (2--1)
observations (Yen et al. 2010).  The inclination corrected $^{12}$CO (2--1)
compact component associated with B 335 is estimated to have a velocity of 160 km s$^{-1}$
with a size scale of 1000 AU, and these high-velocity components are
roughly elongated along the outflow axis.  The $^{12}$CO (2--1) blueshifted
component associated with the L1521F shows similar features in the channel
maps and the PV-diagram (showing a relatively high-velocity component compared
with wide-opening angle outflow), although the velocity range is much
smaller than those observed in B 335.  The centrally condensed
high-velocity components might show a hint of the earliest evolutionary
stage of the high-velocity outflow/jet. 

		In summary, the low-velocity CO emission is originated to the specially extended emission (Figure 2), 
while the high-velocity component is originated to the spatially unresolved compact component (Figure 3). 
Although the extended CO low-velocity emission likely shows the fan-shaped structure (especially at the blueshifted component) 
as seen in protostellar outflow, their velocity structure is not consistent with the 
Hubble-like velocity outflow (i.e., parabolic shaped velocity structure in the PV diagram), which is expected in the wide-angle 
wind model such as observed toward a few Class 0 sources: HH 211, HH 212 and L1448C (Lee et al. 2000 2007a 2007b; Hirano et al. 2010).
The high-velocity components may trace the collimated outflow from the
protostellar core at the earliest evolutionary stage.  Possible
interpretations for the velocity structure of the L1521F outflow will be
further discussed in Section 4.2.2.

\subsection{Outflow Parameters}

The outflow properties obtained from the CO data are calculated following
the standard method by Cabrit \& Bertout (1990). 
The definition of the outflow parameters are same as 
the previous studies of VeLLOs (e.g., Pineda et al. 2001; Dunham et al. 2011), 
so that we can make the comparisons of outflow parameters in following sections.  
As we discussed in the previous section, two velocity components (e.g., the
low-velocity extended component and the higher-velocity compact component)
are suggested.  In order to separately discuss the low-velocity and
higher-velocity components (Section 4.2), the
outflow parameters are estimated according to the following velocity ranges: (i)
low-velocity component (LVC; $V_{\rm{LSR}}$=5.1--6.6 km s$^{-1}$;
${\Delta}V=1.6$ km s$^{-1}$) and (ii) high-velocity component (HVC;
$V_{\rm{LSR}}=2.9-4.5$ km s$^{-1}$; ${\Delta}V=4.2$ km s$^{-1}$) as denoted
in Figure 5.  Note that the systemic velocity of the source
is 6.5 km s$^{-1}$, so our high- and low-velocity definitions are with
respect to that velocity, and only refer to the blueshifted emission. 

Two methods are usually used to calculate the
upper and lower limits to the outflow mass: (i) LTE mass assuming the gas
is optically thin, and (ii) LTE mass with an opacity correction.
Furthermore, we may miss some emission that has been filtered by the
interferometer sampling, therefore, these values are strict lower limits to
the mass. In order to calculate the lower limit of the mass, optically
thin emission was assumed.  Moreover, a significant
amount of redshifted emission is probably missed due to the absorption by
foreground gas (Section 3.3).  Taking account of this, the blueshifted gas was used to
estimate outflow parameters. 

Measuring the flux with a signal level more than \hbox{3$\sigma$} in each channel
\footnote{Images after the primary beam correction were used for the flux measurements}, 
the LTE masses measured in the blue shifted component was estimated to be
\hbox{3.9$\times$10$^{-4}$ M$_{\odot}$} and 5.7${\times}$10$^{-5}$
M$_{\odot}$ for LVC and HVC, respectively.  Here, the excitation temperature of the
\hbox{CO (2--1)} emission was assumed as \hbox{20 K} (e.g., Hogerheijde et
al. 1998; Bourke et al. 2006).  Even when a different excitation temperature is adopted between 
10 --50 K, derived outflow masses change within a factor of 2 (Bourke et al. 2006).  
Line opacity corrections are
also required. The optical depth can be derived as follows; 

		\begin{equation}
			\frac{S_{\nu}(\rm{^{12}CO})}{S_{\nu}(\rm{^{13}CO})} = \frac{[\rm{^{12}CO}]}{[\rm{^{13}CO}]} \left( \frac{1-e^{-{\tau}}}{\tau} \right)
		\end{equation}

where [$^{12}$CO]/[$^{13}$CO] and $S_{\nu}(\rm{^{12}CO})/S_{\nu}(\rm{^{13}CO})$ are the abundance ratio and the intensity ratio 
between $^{12}$CO and $^{13}$CO, respectively. Adopting the abundance ratio [$^{12}$CO]/[$^{13}$CO] 
of $\approx$60 (Wilson \& Rood 1994), and peak intensity ratio of $\approx$7 as $S_{\nu}(\rm{^{12}CO})/S_{\nu}(\rm{^{13}CO})$, 
the optical depth of $\approx$9 was estimated\footnote{Detected $^{13}$CO (2--1)
emission is weak (${\leq}$5$\sigma$) and the $^{13}$CO emission mainly
traces the ambient gas with the missing flux, the opacity estimation may
not be reasonable.  Detailed analysis of the $^{13}$CO will be a future
work with better sensitivity.}. 
The typical optical depth of $^{12}$CO is expected to be 2--5
(Levreault 1988). 
With these opacity corrections, the outflow mass
and related physical parameters will be correspondingly greater than our
lower limits by similar factors.  Assuming that the blueshifted and
redshifted components have a similar outflow mass, the lower limit of the
outflow mass are estimated to be
\hbox{$M_{\rm{CO}}$=2$\times{M_{\rm{CO(blue)}}}$=7.8$\times$10$^{-4}$
M$_{\odot}$} and \hbox{$M_{\rm{CO}}$=1.1$\times$10$^{-4}$ M$_{\odot}$} for
the LVC and HVC, respectively.  Adopting the optical depth of $\sim$9, the
outflow mass for LVC and HVC are estimated to be 7.0$\times$10$^{-3}$
M$_{\odot}$ and 9.9$\times$10$^{-4}$ M$_{\odot}$. 

The outflow momentum ($P_{\rm{flow}}$) and outflow energy ($E_{\rm{flow}}$)  were estimated as 
$P_{\rm{flow}}={\Sigma}M_{\rm{flow}}(j){\times}|v_j-v_{\rm{sys}}|$ and 
$E_{\rm{flow}}=\frac{1}{2}{\Sigma}M_{\rm{flow}}(j){\times}|v_j-v_{\rm{sys}}|^2$, respectively. 
Here $M_{\rm{flow}}$ is the outflow mass in voxel $j$, $v_{\rm{sys}}$ is the 
systemic velocity of 6.5 km s$^{-1}$, and $v_j$ is the velocity of voxel $j$. 
The outflow characteristic velocity $v_{\rm{flow}}$ is defined as $P_{\rm{flow}}/M_{\rm{flow}}$. 
The outflow size, $R_{\rm{flow}}$, was measured from the integrated intensity map shown in Figure 2, 
where the emission signal to noise ratio of more than 3. Measured size was not deconvolved size. 
The outflow characteristic velocity and the outflow size can be corrected for the inclination
angle $i$ with \hbox{$v_{\rm{flow}}=v_{\rm{obs}}{\times}[1/{\sin{i}}$]} and
\hbox{$R_{\rm{flow}}=R_{\rm{obs}}{\times}[1/{\cos{i}}$]}, respectively. 
In this case, projected characteristic outflow velocity and outflow size are 
defined as $v_{\rm{obs}}$ and $R_{\rm{obs}}$, respectively. 
The inclination angle of L1521F-IRS (envelope structure) was estimated to
be \hbox{50$^{\circ}$--70$^{\circ}$} from the SED and infrared image by
Terebey et al. (2009).  This corresponds to an outflow inclination angle
of between 20$^{\circ}$--40$^{\circ}$.  For the outflow estimation, the
mid value of 30$^{\circ}$ is assumed.  
Varying the inclination angle between 20$^{\circ}$ and 40$^{\circ}$ changes 
the outflow size and outflow velocity factor of 1.2 and 1.9, respectively. 
The dynamical time of the outflow is calculated to be $t_d=R_{\rm{flow}}/v_{\rm{flow}}$ using the characteristic outflow velocity. 
Finally, momentum rate (outflow force), mechanical luminosity, and mass loss rate are calculated to be \hbox{($F_{\rm{obs}}=P/t_d$)}, 
\hbox{($L_m=M_{\rm{CO}}{\times}v_{\rm{flow}}^3/2R$)}, \hbox{($\dot{M}_{\rm{out}}=M_{\rm{CO}}/t_d$)}, respectively. 

Note that the upper limit and inclination corrected outflow parameters are
used in the following subsections.  This is because the CO (2--1) emission
is most likely optically thick in the molecular outflow (e.g., Leveault
1988 as well as our estimations; See section 3.3.2).  Moreover, in order to
include the redshifted outflow, which is not detected (most
probably due to the self-absorption as discussed in Section 3.3), the
outflow parameters described in the following Section are simply multiply factor two of 
the blueshifted outflow parameters.

\section{DISCUSSION}		

\subsection{Origin of the SMA 1.3 mm Continuum Emission}

		Our SMA observations detected 1.3 mm continuum emission
associated with the Spitzer source, L1521F-IRS at an angular resolution of
$4.8''{\times}4.4''$ although its peak position is shifted from L1521F-IRS. 
Recent PdBI observations at a much higher angular resolution by Maury
et al. (2010), $0.49''{\times}0.27''$, detected a weak and compact
continuum source embedded within the extended continuum source detected with
the SMA.  In this Section, we will discuss the origin of the dust
continuum emission. 
		
		Crapsi et al. (2004) fitted the 1.2 mm continuum results
obtained from a single-dish telescope with a flat-top core density
structure using the semi analytic model of
\hbox{$n(r)=n_{0}/(1+(r/r_0)^{\alpha}$)} (Tafalla et al. 2002) with
\hbox{$n_{0}=10^6$ cm$^{-3}$}, \hbox{$r_0=$2800 AU}, and \hbox{$\alpha$=2},
which is consistent with an isothermal sphere.  
In order to investigate the origin of the observed SMA 1.3 mm continuum
emission on smaller scales, simple power-law models are compared to the
data.  Specifically, the (i) inside-out collapse model --
${\rho}(r){\propto}r^{-1.5}$ by Shu (1977), and (ii) isothermal sphere
model -- ${\rho}(r){\propto}r^{-2}$ by Larson (1969) and Penston (1969), are
computed and directly compared to the SMA data using the SMA uv-coverage
and imaged with the same tapering parameter we used in \hbox{Figure 1}. 
Here, we adopted the gas temperature of 10 K (e.g., Myers \& Benson 1983). 
We assumed that the parental core has a diameter of 4200 AU and the total
flux is 5.8 Jy at 1.3 mm based on 1.2 mm single-dish results by Kauffman et
al. (2008). The total and peak fluxes are measured from the simulated image
using 2D Gaussian fitting, finding (i) total flux density of \hbox{$\sim$11
mJy} for the inside-out collapse model, and (ii) total flux density of
\hbox{$\sim$50 mJy} for the isothermal sphere core model. This shows that
the the observed flux density, 11 mJy, measured from the two dimensional
Gaussian fitting is comparable to or smaller than that expected to arise
from inner regions of a molecular envelope, suggesting that the observed
emission is unlikely arising from an additional component, such as a disk,
in addition to an envelope.

		The 1.3 mm continuum emission detected with the PdBI is
probably arising from the most inner part of the gas envelope because 
the PdBI continuum position is coincide with {\it Spitzer} source position (L1521F-IRS).  
The positional shift of the peak position between the SMA
and PdBI continuum results is not clear because the extended structure
detected with the SMA have not been spatially resolved and the visibility
amplitude plot does not indicate clear structure. 

One possible interpretation to explain the positional shift is the asymmetrical 
nebula structure originated to the outflow inclination. 
As we seen in Figure 2(a), the infrared reflection nebula is brighter at west, while 
less brighter at east. This is likely due to that the western lobe is pointing toward us, 
and the emission from the eastern lobe is partially absorbed by the foreground dense 
gas. If similar situation occur in the 1.3 mm continuum emission, the emission peak 
could be shifted toward the western lobe.

Our SMA observations and the PdBI observations by Maury et al. (2010) have different mass sensitivities and uv-coverage. 
The 1.3 mm peak flux reported from the PdBI observations, \hbox{1 mJy beam$^{-1}$} (with the PdBI beam size of \hbox{$0''.49{\times}0''.27$}), 
is similar to the noise level of the SMA image, \hbox{1.3 mJy beam$^{-1}$} (with the SMA beam size of \hbox{$4''.81{\times}4''.43$}). 
This huge beam surface area difference between the PdBI and SMA (factor of $\sim$160) suggests that the present SMA observations 
are not able to detect the compact continuum peak detected with the PdBI. 
Furthermore, the PdBI observations cover the baseline ranging from \hbox{24 m} to \hbox{760 m}, 
while our SMA observations cover the base line range of \hbox{12 m} to \hbox{69 m}.
Even with the simple $uv$-length range comparison, the SMA observations are at least twice as sensitive to the extended emission compared with PdBI observations, 
but PdBI observations are more sensitive for resolving the compact structure. 
				
		Imaging simulations and observational sensitivity
comparisons suggest the 1.3 mm continuum emission detected with the SMA most likely
traces the thermal dust emission from the inner envelope rather than
emission from circumstellar disk, while the 1.3 mm continuum emission
detected with the PdBI as well as $Spitzer$ sources trace the warm dust
originated from the inner envelope/circumstellar disk.

\subsection{Nature of L1521F}

The main purpose for studying L1521F is to understand nature of this
source as well as the nature of VeLLOs.  L1521F (MC 27) was originally noticed
as a starless condensation with a high central density of $\sim$10$^{6}$
cm$^{-3}$ (Onishi et al. 1998).  Infalling asymmetry spectra observed in
the HCO$^{+}$ (4--3/3--2) emissions suggest infalling gas motion with the
size scale of \hbox{2000-3000 AU} (Onishi et al. 1999).  Molecule depletion
observed in CCS and the enhanced deuterium fraction imply the chemically
evolved starless core (Shinnaga et al. 2004; Crapsi et al. 2004), which may be 
about to form a star. The {\it Spitzer} observations discovered a
reflection nebula and a point source associated with L1521F (Bourke et al.
2006).  As noted in Terebey et al. (2009) and Saigo et al. (2011), \hbox{L
1521-IRS} is not likely FHSC since the near to mid
infrared source (i.e., Spiter/IRAC bands) was clearly detected.  In
addition, detection of the 50 AU scale compact continuum source at
\hbox{1.3 mm} may suggest the presence of a circumstellar disk (Maury et al.
2010).  Detection of compact high-velocity gas in the \hbox{CO (2--1)} line
also supports the presence of a molecular outflow.  These observational
facts suggest that \hbox{L1521F-IRS} is already in the protostellar phase
(i.e., after second collapse phase) with the very low internal luminosity of $<$0.07 L$_{\odot}$ 
(Bourke et al. 2006; Dunham et al. 2008; Terebey et al. 2009).

\subsubsection{Evolutionary Status}

Here, we assume a standard accretion rate at the main accretion phase,
\hbox{1.6$\times$10$^{-6}$ M$_{\odot}$ yr$^{-1}$}, derived from $c_s/G^3$
(e.g., Shu et al. 1977) at T$\approx$10 K (e.g., Myers \& Benson 1983). 
Here, c$_s$ is the sound speed and G is the gravitational constant. 
This accretion rate is similar to those directly measured in low-mass star
forming regions (e.g., Ohashi et al. 1997; Saito et al. 2001).  The current
stellar mass can be estimated from
$L_{\rm{acc}}=GM_{\ast}{\dot{M}_{\rm{acc}}}/R_{\ast}$.  Considering the
internal luminosity of L1521F of 0.034--0.07 L$_{\odot}$ (Bourke et al.\ 2006; Tereby et al. 2009) 
is mainly due to the accretion luminosity
$L_{\rm{acc}}({\approx}L_{\rm{int}})$, and assuming the stellar radius is
\hbox{$R_{\ast}{\sim}3R_{\odot}$} (Palla 2002), and the standard accretion
rate, the stellar mass is estimated to be \hbox{0.002--0.004 M$_{\odot}$}, or \hbox{2--4 M$_J$}. 

The estimated mass of the central protostar, assuming the standard accretion rate is a factor of 2.5--5 smaller than the minimum mass of FHSC or a second core,
0.01 M$_{\odot}$, derived from the theoretical calculations (e.g., Masunaga\& Inutsuka 1998; Bete 2011; Machida et al. 2011).  
In order for the current stellar mass to be 0.01 M$_{\odot}$ at the observed L$_{\rm{int}}$, the mass accretion rate should be 
at least four times smaller, i.e.,$\dot{M}_{\rm{acc}}{\sim}$4$\times$10$^{-7}$.  
Such a small mass accretion rate is not expected in standard star formation
theories, whereas it can be expected in a scenario of non-steady accretion
(e.g., Hartmann \& Kenyon 1996; Vorobyov \& Basu 2005, 2006; Machida et al.
2011). Recent Magnetro hydrodynamic simulations (MHD simulations) by
Machida et al. (2011) suggest that the mass accretion rate suddenly
decreases to 10$^{-5}$ M$_{\odot}$ yr$^{-1}$ right after a central
protostar is formed, and the accretion rate gradually decreases afterward.
We note that such a small mass accretion rate would be inconsistent with a
relatively larger mass outflow rate of L1521F-IRS because
$\dot{M_{\rm{w}}}{\approx}(0.1-0.5){\dot{M}_{\rm{acc}}}$ (e.g., Hartigan et
al. 1995), where $\dot{M}_{\rm{w}}$ is the mass loss rate.  When we assume
that 10\% of the accreting material is ejected as an outflow, the mass
accretion rate is estimated to be $\dot{M}_{\rm{acc}}$=1.5$\times$10$^{-6}$
--9.7$\times$10$^{-5}$ M$_{\odot}$ yr$^{-1}$.  
Here, the mass accretion rate were calculated
based on the mass outflow rate listed in Table 2 as
$\dot{M}_{\rm{acc}}=\dot{M}_{\rm{w}}/0.1=2{\times}
(\dot{M}_{\rm{out(LVC)}}+\dot{M}_{\rm{out(HVC)}})/0.1$. 
This mass outflow rate, however, is an outflow rate averaged over the period of the dynamical
time scale of the outflow, meaning that the mass infall rate estimated from
the mass outflow rate based on
$\dot{M_{\rm{w}}}{\approx}0.1{\dot{M}_{\rm{acc}}}$ is also an averaged
value over the dynamical time scale.  When non-steady accretion is
considered, the current accretion rate can be smaller than the averaged
mass accretion rate. 
Dunham et al. (2011) made comparisons of outflow parameters measured toward
VeLLOs that have been observed at high angular resolution, which includes
candidate FHSCs, very young protostellar cores, and
candidate proto-brown dwarfs.  The mass outflow rate toward them,
$\dot{M}_{\rm{out}}=M_{\rm{flow}}/{\tau}_d$, falls in the range
$6.0\times10^{-9} - 1.1\times10^{-6}$ M$_{\odot}$ yr$^{-1}$. 
Estimated outflow rate toward L1521F is comparable or slightly higher than those observed by Dunham et al. (2011). 

As discussed above, the expected time averaged mass accretion rate suggests
that half of the VeLLOs reported by Dunham et al. (2011), i.e., 
L1448-IRAS 2E, CB17 MMS, L1148-IRS, and L673-7-IRS, have
comparable or larger mass accretion rates compared to those expected in the
standard low-mass star formation scenario. This suggests that central
stellar mass larger than 0.01 M$_{\odot}$ should already be present, in the
case of steady accretion.  Time variable mass accretion processes (i.e.,
lower mass accretion rate at present), which is also suggested for L1521F,
could be responsible for the observed low internal luminosities for some
VeLLOs.

\subsubsection{Outflow Morphology and Velocity}

Three-dimensional MHD simulations have been performed on molecular outflows
driven by FHSCs and second cores (i.e., protostars; Machida
et al. 2008).  These simulations predict the presence of a low-velocity
flow \hbox{(5 km s$^{-1}$)} with a wide opening angle, driven from FHSC \hbox{($n_c>10^{12}$ cm$^{-3}$)}, while high velocity flows
\hbox{(30 km s$^{-1}$)} with good collimation, are expected when the
central protostar (2nd core; \hbox{$n_c>10^{21}$ cm$^{-3}$}) is formed. 
Our CO (2--1) outflow observations spatially resolved the less collimated
morphology in the low-velocity component of the L1521F outflow
(${v}_{\rm{flow}}=1.5$ km s$^{-1}$), while the high-velocity outflow
(${v}_{\rm{flow}}=4.0$ km s$^{-1}$) is extremely compact and associated only with the central region around L
1521-IRS.  An interesting characteristic of the extended low-velocity
outflow is that it does not follow a Hubble-like velocity structure, which
is predicted by the wind-driven model as discussed in \hbox{Section 3.3.1}.  
These results suggest that the nature of the low-velocity component
detected from \hbox{L1521F-IRS} may not be typical of outflows/jets
associated with the Class 0/I sources. However, it could be related to the
outflow driven from FHSC since they are expected to be
of very low-velocity and less collimated, as shown by MHD simulations
(Machida et al. 2008). 

In reality the outflow from FHSC may have more variety
than the less collimated morphology predicted by simulations (Machida et
al. 2006; 2008).  A very recent smoothed particle magnetohydrodynamics
(SPMHD) simulations by Price et al. (2012) demonstrated that collimated
jets (opening angles $\leq$10$^{\circ}$) may be present during the collapse
of molecular cloud cores to form FHSCs in low-mass star
formation.  Collimated molecular outflow have been observed from candidate
FHSCs such as L1448 IRAS 2E (Chen et al.  2010) and Per-Bolo 58 (Enoch et al.  2010; Dunham et al.
2011). 

The detected high-velocity CO (2-1) emission likely traces a collimated
molecular outflow, although we have not completely spatially resolved the
structure.  The measured characteristic outflow velocity after the inclination correction
was estimated to be 4.0 km s$^{-1}$ (or the maximum outflow velocity of 7.2 km s$^{-1}$). 
This velocity is roughly consistent with the velocity expected of collimated jets originating from FHSCs, 
2--7 km s$^{-1}$ (Price et al. 2012).  However, in L1521F
the compact CO (2--1) emission is unlikely to originate from a FHSC, as it is associated with the protostar.  

The dynamical time scale of the low- and high-velocity components
associated with L1521F are estimated to be 3800 yr and 1100 yr,
respectively.  If we assume that the times for the low- and high-velocity
outflows reflect the age of first core and second core, respectively, then
the age of the second core should also be about 1100 yr.  Further, about
2700 yr after the first core was formed, the second collapse occurred.  The
estimated life time of FHSC is roughly consistent with
those estimated from MHD simulations (Machida et al. 2011).  Moreover, the
estimated mass of the second core is approximately 0.01 M$_{\odot}$ about
1000 yrs after the second collapse. This is also consistent with the mass
of L1521F-IRS estimated from the internal luminosity in the case of
non-steady accretion. 

Observations with improved angular resolution and better sensitivity are
needed for more detailed comparisons with MHD simulations, and to study the
nature of the extended outflow.  This will be an important study using 
ALMA. 

\subsubsection{Comparisons of Outflow Characteristics}

	Figure 6 compares the outflow parameters of L1521F (this work)
with previously studied outflows associated with low- to high-mass star
forming regions (Wu et al. 2004), and outflows associated with VeLLOs
(referred from Table 4 in Dunham et al. 2011).  The measured outflow mass,
size, and outflow force of VeLLOs including L1521F (as well as all the
other parameters listed in Table 2) show a small range of values compared
to outflows from low to high-mass star forming regions,
and are located at the lower end of the range of values seen in all
outflows.  On the other hand, the derived L1521F outflow parameters share
similar characteristics with outflow parameters derived from other VeLLOs
as summarized in Dunham et al. (2011). 

	Outflow masses and sizes are integrated values over the outflow
evolution.  As clearly seen in Figure 6 (a) and (b), the outflow mass and
size reported for VeLLOs including L1521F show orders of magnitude smaller
values compared with those of the main population of the previously studied
outflows.  This may suggest that the outflows associated with VeLLOs are
not well developed yet.  The dynamical time scale of the outflow could also 
support this argument: the outflow dynamical time scales derived in VeLLOs are 700
to 61000 yrs, while those derived from previous outflow studies summarized
in Wu et al. (2004) ranges 2.0$\times$10$^3$--5.0$\times$10$^5$ yrs
Note that the outflow dynamical time scale can only tell us lower limits to the real outflow age. 
Especially if the source become distant, the size underestimation effect become more significant 
due to the sensitivity limit. However, considering with the distance effect, we still see that the 
dynamical time scales associated with non-VeLLOs are orders of magnitude shorter in average 
compared to those in more massive outflows. This clearly implies outflow character differences 
between them, and  
the results imply that L
1521F as well as some of VeLLOs are probably in the earliest evolutionary
stage of the protostellar phase ($<$10$^4$ yr, which is yonger than Class
0).  Alternatively, it could be that emission from these outflows is hard
to detect at larger distances from their driving sources, as the emission
is often faint, and may be below the sensitivity limits of current
observations.

	The dashed square in Figure 6(b) and (c) shows the area where
outflows associated with embedded sources in the Taurus star forming region
are located (Hogerheijde et al. 1998).  L1521F has the lowest bolometric
luminosity (i.e., internal luminosity) and the smallest outflow size among
the Taurus protostellar candidates.  Nevertheless, the estimated outflow
parameters such as outflow mass and force, as well as
outflow momentum, energy, and mechanical force are comparable to the
previously studied low-mass outflow cases.  For example, the outflow force
derived in embedded protostars in Taurus is
\hbox{1.9$\times$10$^{-7}$--2.2$\times$10$^{-3}$} \hbox{M$_{\odot}$ km
s$^{-1}$ yr$^{-1}$} (e.g., Hogerheijde et al. 1998), while the value for
the L1521F outflow is $7.4{\times}10^{-7}$--$6.6{\times}10^{-6}$ M$_{\odot}$ km s$^{-1}$
yr$^{-1}$ (within the dashed square in Figure 6(c)).  The outflow
parameters reflecting outflow activity such as the outflow force are
comparable to some outflows in Taurus.  A possible explanation is that L
1521F could be a low-mass protostar similar to other protostars in Taurus.
The averaged outflow force (as well as the averaged mass outflow rate and
outflow energy) is larger than other outflows in Taurus, while the
relatively smaller outflow size (as well as short outflow dynamical time
scale) suggest that the L1521F-IRS is in an earlier evolutionary phase than
other embedded sources in Taurus.

	In summary, the L1521F outflow is one of the smallest in size 
and youngest in dynamical time among nearby low-mass protostellar cores as same as other VeLLOs, 
suggesting that L1521F-IRS is in a very early evolutionary phase
($<$10$^4$ yr).  The estimated outflow force, energy, and mass loss rate
suggests that the time averaged star formation activity of L1521F-IRS may
be similar to other low-mass protostars in Taurus.  The extremely smaller
internal luminosity of L1521F-IRS might suggest that the mass accretion
rate at present is much smaller than the time-averaged value.

\section{SUMMARY AND FUTURE PROSPECTS}

We have performed CO(2--1) and 1.3 mm continuum observations toward the
very low-luminosity object, L1521F-IRS, with the SMA.  The main results
are summarized as follows:

	\begin{itemize}

	\item		We have spatially resolved a 
molecular outflow associated with \hbox{L1521F-IRS} in CO (2--1)
emission for the first time.  The blueshifted and redshifted lobes are
aligned along the east and west side of \hbox{L1521F-IRS} with a lobe size
of \hbox{$\approx$1000 AU}. The estimated outflow mass, the maximum outflow
velocity, and the outflow force are \hbox{(9.0-80)${\times}$10$^{-4}$
M$_{\odot}$}, 7.2 km s$^{-1}$, and \hbox{(7.4-66)$\times$10$^{-7}$ M$_{\odot}$
km s$^{-1}$ yr$^{-1}$}, respectively. 
			
	\item		The estimated outflow size, mass, and energy are
located at the lower end of parameters derived from previously
studied outflows associated with low- to high-mass star forming regions,
but share similar characteristics to outflows associated with VeLLOs. The
CO emission suggests a low-velocity less collimated component (1.5 km
s$^{-1}$/1200 AU) and a high-velocity compact component (4.0 km
s$^{-2}$/920 AU).  These velocity structures are not consistent with those
expected in the jet driven or wind driven outflows, maybe suggesting the
presence of the outflow remnant from the FHSC as well as
the undeveloped outflow from the protostar. 
			 
	\item		The previous detection of an infrared source and
compact dust continuum emission suggests the presence of the protostar.
Its low bolometric luminosity ($\approx$0.05 L$_{\odot}$) and small outflow
lobe size, among the smallest of all previously studied outflows,
suggest that L1521F-IRS is at least one order of magnitude younger
($<10^{4}$ yr) than Class 0 protostars and contains (at present) a
substellar object at the center. 
A comparison between mass outflow rate and minimum requirement for 
the theoretically predicted initial stellar/core suggests a mass, non-steady accretion model  
could explain the observational results without contradiction. i.e., the current 
accretion rate can be smaller than the averaged mass accretion rate.

	\item		Our 1.3 mm continuum observations with the SMA have
detected an extended thermal dust component with a mass of 0.047--0.084
M$_{\odot}$.  Our imaging simulations suggest that this emission originates
from the inner envelope rather than from a circumstellar disk. 

	\end{itemize}

\acknowledgments
We acknowledge the staff at the Submillimeter Array for assistance with
operations. The SMT is operated by the Arizona Radio
Observatory (ARO), steward Observatory, University of Arizona.  S. T.
acknowledges S.-W. Yen for helping the SMT $^{12}$CO (2--1) data
inspection, and M. Machida and N. Hirano for fruitful scientific comments.
S. T. is financially supported by a postdoctoral fellowship at the
Institute of Astronomy and Astrophysics, Academia Sinica, Taiwan.

\clearpage

\begin{figure}
\epsscale{0.7}
\plotone{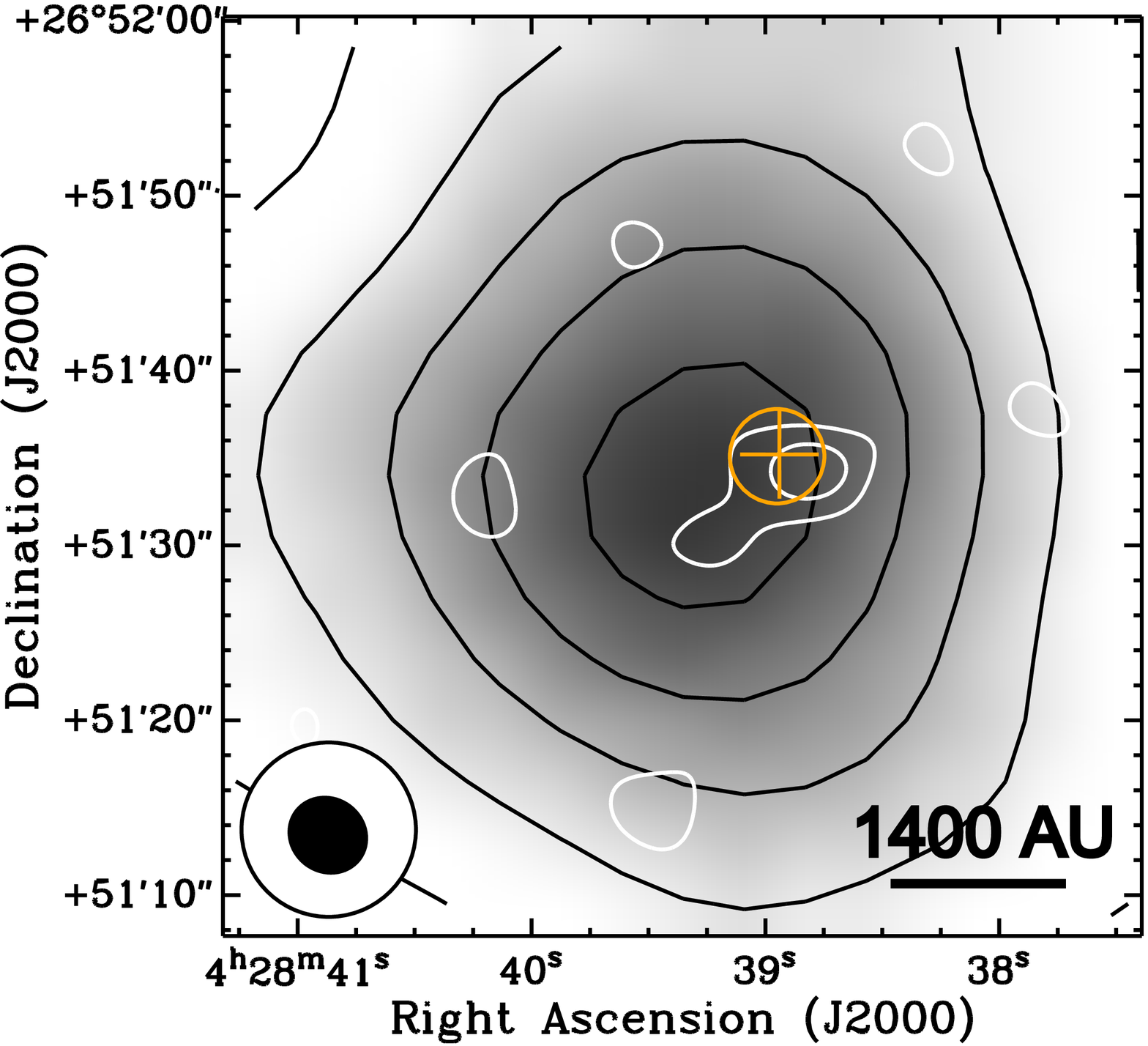}
\caption{The 1.3 mm continuum image from SMA observations in white contours
superposed on the 1.2 mm continuum image taken with the IRAM 30m/MAMBO-2 by
Kauffmann et al. (2008) in grey scale and black contours. The contour
levels of the SMA continuum emission start at ${\pm}2{\sigma}$ with a
interval of 2${\sigma}$ (1${\sigma}$=1.3 mJy beam$^{-1}$).  The contour
levels of the 1.2 mm image taken with the IRAM 30m/MAMBO-2 are 29.5, 37.5,
45.5, 53.5, and 61.5 mJy beam$^{-1}$, respectively.  The cross and circle
show the positions of the PdBI 1.3 mm and Spitzer sources (L1521F-IRS),
referred from Maury et al. (2010) and Bourke et al. (2006), respectively.  
Open and filled ellipses in the bottom left corner shows the beam sizes of
IRAM 30m/MAMBO-2 and SMA, respectively. 
\label{f1}}
\end{figure}

\begin{figure}
\epsscale{0.8}
\plotone{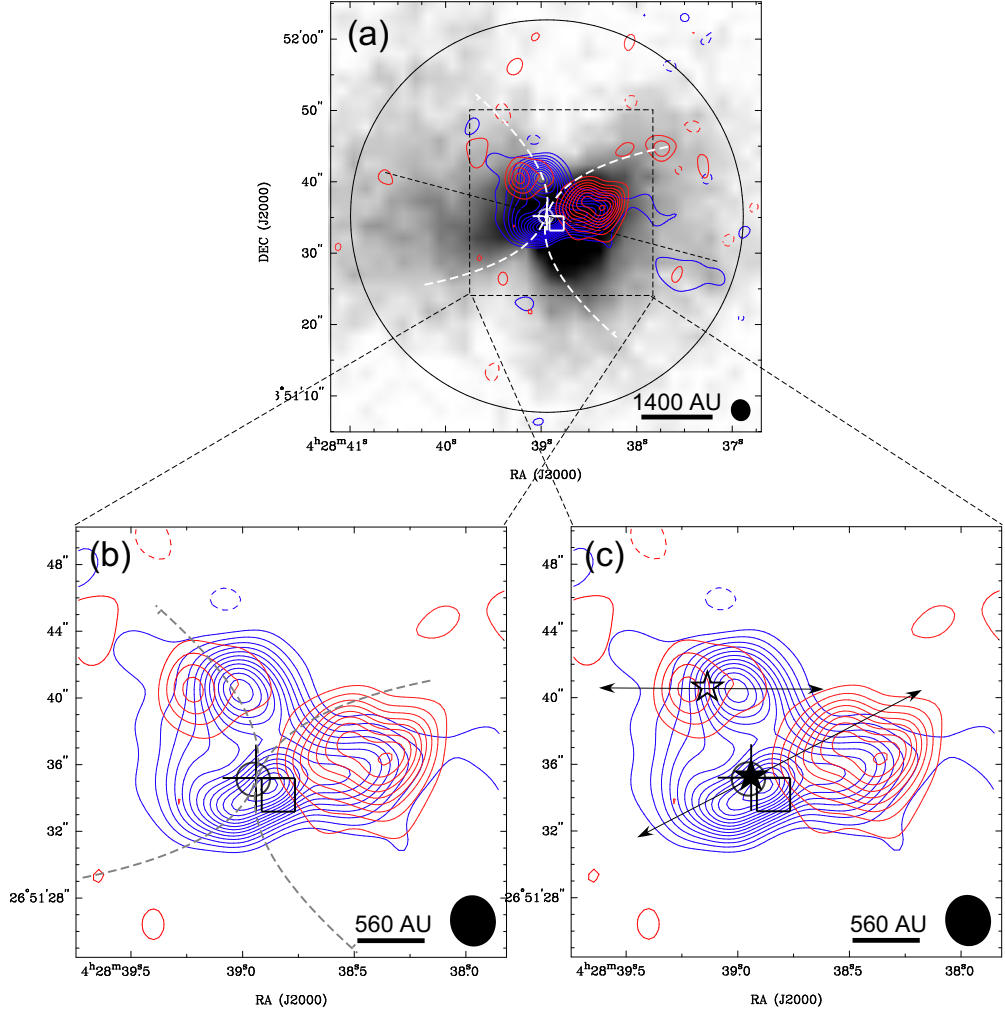}
\caption{
(a) Blueshifted and redshifted velocity components of the CO (2--1) emission in
contours taken with the SMA superposed on the 4.5 $\mu$m image obtained with
{\it Spitzer}/IRAC (Bourke et al. 2006).  The large circle represent the
FWHM size of the SMA primary beam, while the filled ellipse at bottom right
shows the synthesized beam size.
(b) Zoomed in image of the CO (2--1) blueshifted and redshifted components.  The
velocity ranges of the blueshifted and redshifted emission are
V$_{\rm{LSR}}$=2.9--6.1 km s$^{-1}$ and V$_{\rm{LSR}}$=8.2--10.4 km
s$^{-1}$, respectively.  The contour lines start at ${\pm}3{\sigma}$
levels, with intervals of 3$\sigma$ (1${\sigma}_{\rm{(blue)}}=0.35$ Jy
beam$^{-1}$ km s$^{-1}$; 1${\sigma}_{\rm{(red)}}=0.24$ Jy beam$^{-1}$ km
s$^{-1}$).  Negative contours are represented by dashed lines.  The peak
positions of the PdBI 1.3 mm continuum source (Maury et al. 2010), infrared
(IRAC) source (Bourke et al. 2006), and SMA 1.3 mm continuum source (this
work) are denoted in the cross, open circle, and square, respectively.
The green and grey dotted lines show the guide lines for the parabolic
shaped molecular outflow originated from the wide angle wind model as
discussed in Section 3.4. 
(c) The schematic picture of possible multiple outflows. Open and filled
stars present the position of L1521F-IRS and L1521F-NE (if exists).  
\label{f2}}
\end{figure}

\begin{figure}
\epsscale{0.9}
\plotone{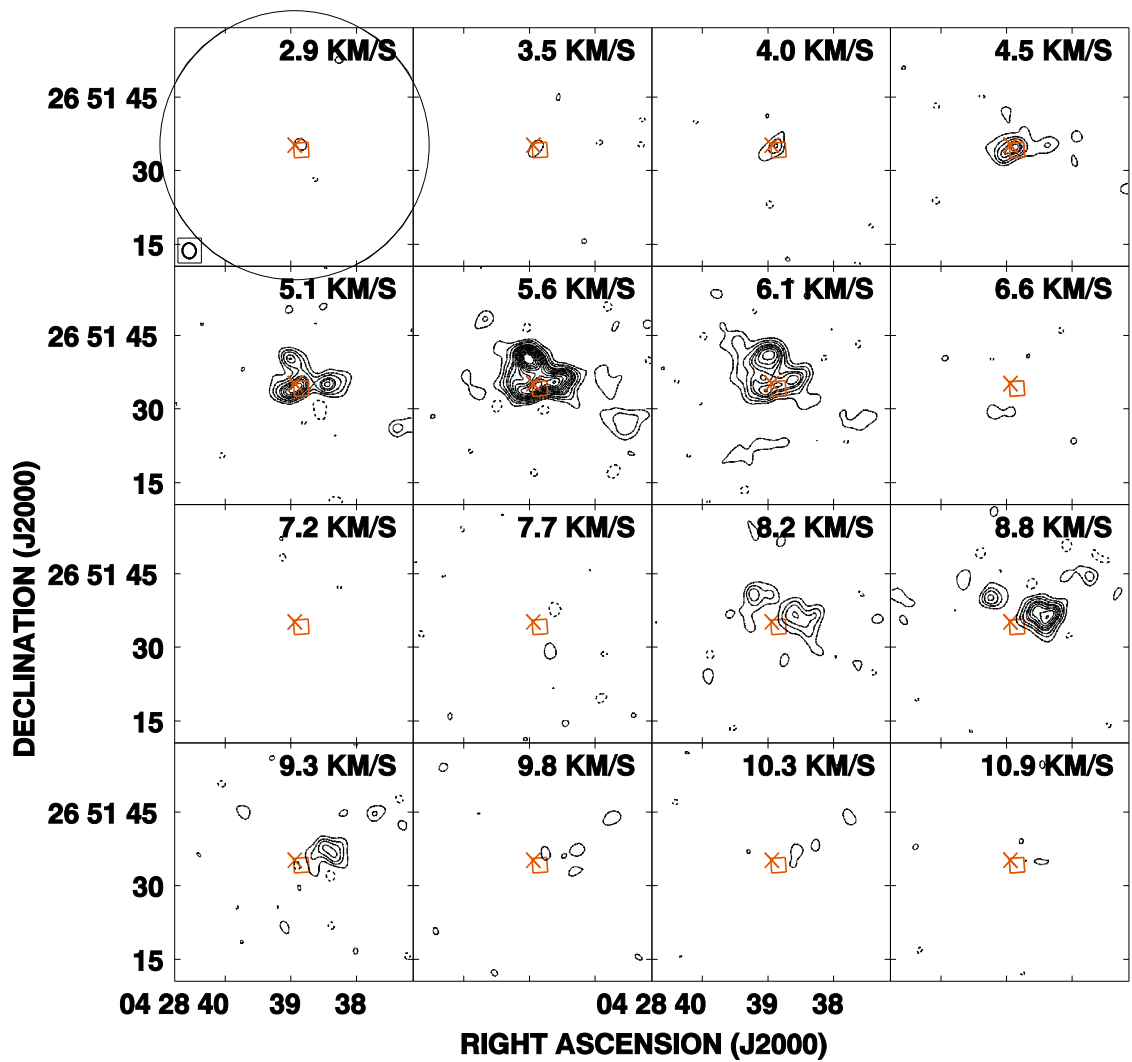}
\caption{Velocity channel maps of the $^{12}$CO (2--1) emission taken with
the SMA.  The central LSR velocity in units of km s$^{-1}$ is noted at the
upper right corner of each panel.  The contour levels start at 3${\sigma}$
levels, with intervals of 3${\sigma}$ (1${\sigma}$=0.13 Jy beam$^{-1}$).
Crosses and square in the maps show the peak positions of the PdBI 1.3 mm
(as well as infrared source detected with {\it Spitzer/IRAC}) and SMA 1.3
mm source, respectively.  The large circle in the upper-left panel shows
the SMA primary beam size, while the open ellipse in the bottom left corner
shows the SMA synthesized beam, respectively.  Negative contours are shown
as black dashed lines. 
\label{f3}}
\end{figure}

\clearpage
\begin{figure}
\epsscale{0.6}
\plotone{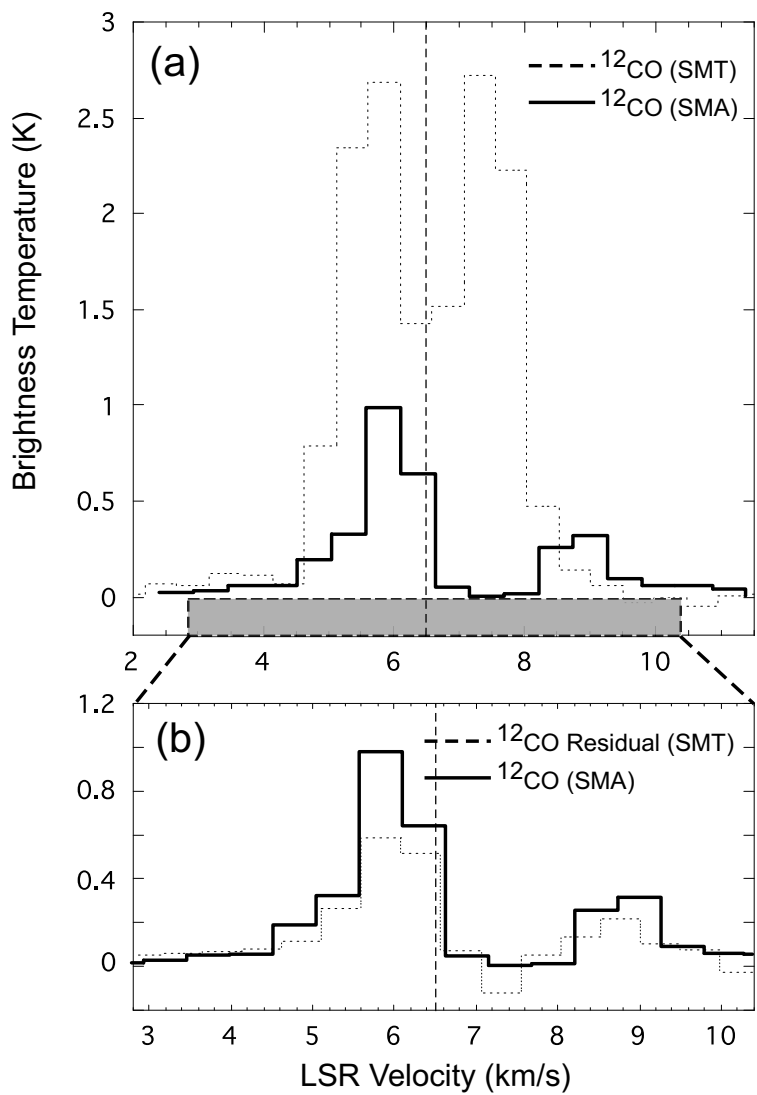}
\caption{
(a) Comparison of the SMT and SMA $^{12}$CO (2--1) spectra at the position
of L1521F-IRS.  The SMA data are convolved to the SMT 32$''$ beam, 
(b) $^{12}$CO (2--1) spectrum taken with the SMA compared with the $^{12}$CO
residual spectrum. The residual spectrum is derived by subtracting from the central 
spectrum the mean spectrum from positions 30$''$ away. Dashed lines show 
the system velocity of L1521F of 6.5 km s$^{-1}$.  
\label{f4}}
\end{figure}

\begin{figure}
\epsscale{0.9}
\plotone{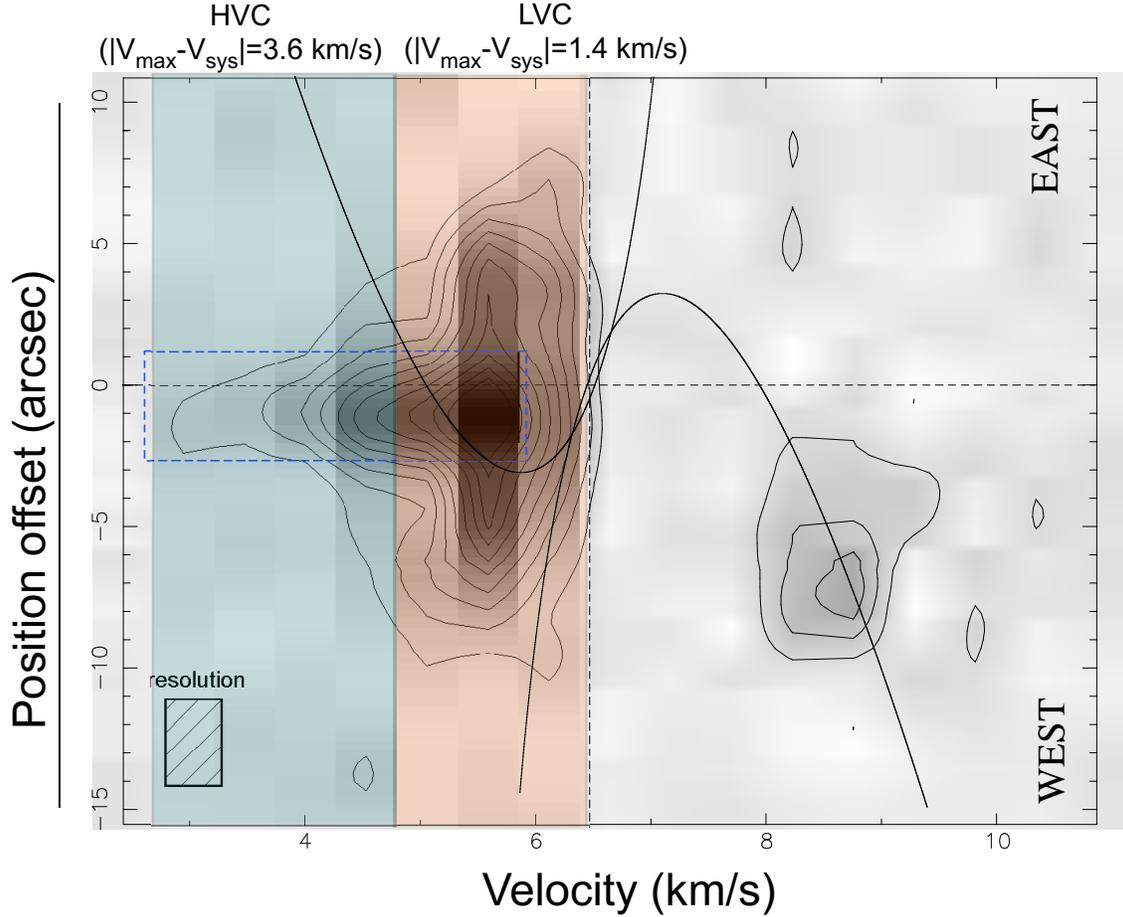}
\caption{Position-velocity diagram of $^{12}$CO (2--1) cut along the outflow axis (P. A.
=75$^{\circ}$).  The contour level starts at 3 $\sigma$ with intervals of
3 $\sigma$ (1 $\sigma$=0.13 Jy beam$^{-1}$ km s${-1}$).  Dashed lines in
the diagram show the systemic velocity (6.5 km s$^{-1}$) and the protostar
position.  The solid black curve is the wide-angle wind model curve produced
by setting $i=60^{\circ}$, $C$=0.022 arcsec$^{-1}$, $v_0=0.05$ km s$^{-1}$.  The
filled square at the bottom left corner shows the spatial and velocity
resolution.  The red box and blue box show the velocity ranges of the
low-velocity component ($|v_{\rm{max}}-v_{\rm{sys}}|=1.4$ km s$^{-1}$) and
the high-velocity component ($|v_{\rm{max}}-v_{\rm{sys}}|=3.6$ km
s$^{-1}$), respectively. 
\label{f5}}
\end{figure}

\begin{figure} 
\epsscale{1.0} 
\plotone{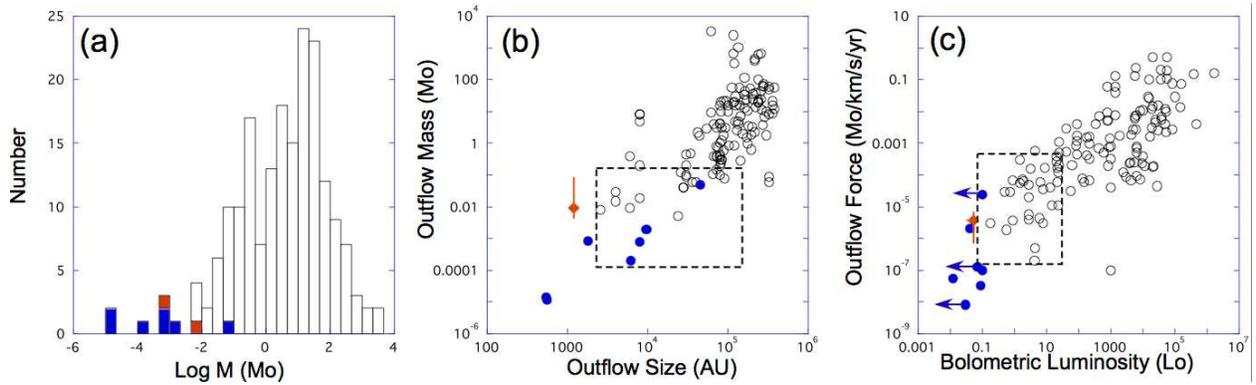} 
\caption{
(a) Outflow mass histogram. The lower limit (optically thin condition) and upper limit ($\tau$=9) 
estimated  in L1521F was denoted by orange color, (b) outflow mass plotted as a function of
outflow size, and (c) outflow force plotted as a function of bolometric
luminosity.  Orange symbols/histogram denote the outflow from L1521F (this
work). Open symbols/histograms show parameters reported in previous
low to high-mass outflow studies by Wu et al. (2004), while blue
symbols/histograms show parameters reported for outflows associated
with VeLLOs by Dunham et al. (2011). The values presented for 129
outflows extracted from Wu et al. (2004) are obtained from online tables that list outflow size, 
mass, force, and bolometric luminosity of the central source. 
The outflow parameters range derived in Taurus by Hogerheide et al. (1998) was denoted as dashed squares 
in figure (b) and (c).
\label{f6}}
\end{figure}

\clearpage
\begin{deluxetable}{lc}
\tabletypesize{\scriptsize}
\tablecaption{SMA Observational Parameters\label{tbl1}}
\tablewidth{0pt}
\tablehead{
\colhead{Parameter} & \colhead{Value} \\
}
\startdata
Observing Date (UT)													& Jan. 3, 2007 \\
Reference position (J2000.0)			                            & $\alpha$=4$^h$28$^m$38.95$^s$, $\delta$=+26$^{\circ}$51'35.10'' \\
Configurations	  					                            	& compact \\
Primary beam HPBW [arcsec]	  			                  			& 55 \\
Synthesized Beam HPBW of 1.3 mm continuum emission [arcsec]	  	    & 4.8$\times$4.4 (53)  \\ 
Synthesized Beam HPBW of CO (2--1) emission [arcsec]	  	            & 3.1$\times$2.7 (9.1)  \\ 
Equivalent frequency [GHz]                                          & 214.534 \\
Bandwidth [GHz]	                                                    & 4.0 \\ 
Velocity resolution of CO (2--1) emission [km s$^{-1}$]				& 0.53 \\
Projected base line range of continuum / CO (2-1) [k$\lambda$]      & 9.1--53 / 9.5--53 \\
Maximum detectable structure of continuum/CO (2-1) [arcsec]			& 18 / 17 \\
Gain calibrators (flux of Gain calibrators in Jy)                   & 3C 111 (3.0 Jy) \\
Bandpass calibrator	                                                & 3C 279 \\
Primary flux calibrators                                           	& Uranus \\  
RMS noise level of continuum emission [mJy beam$^{-1}$]             & 1.3 \\
RMS noise level of CO (2--1) emission [Jy beam$^{-1}$]             	& 0.13 \\
\enddata
\tablenotetext{a}{Our observations were insensitive to more extended emission than this size-scale structure at the 10\% level (Wilner \& Welch 1994).}
\end{deluxetable}

\begin{deluxetable}{lllll}
\tabletypesize{\scriptsize}
\tablecaption{Outflow Parameters Measured from the Blueshifted Emission \tablenotemark{d}}
\tablewidth{0pt}
\tablehead{
\colhead{Parameters} & \multicolumn{2}{c}{LVC}  & \multicolumn{2}{c}{HVC} \\ \cline{2-5}
\colhead{} & \colhead{} & \colhead{} & \colhead{} & \colhead{} \\
\colhead{} & \colhead{Uncorrected\tablenotemark{a}} & \colhead{Corrected\tablenotemark{b}} & \colhead{Uncorrected\tablenotemark{a}} & \colhead{Corrected\tablenotemark{b}} \\
}
\startdata
Mass ($M_{\odot}$)\tablenotemark{c}						&	(3.9--35)e-04	&	(3.9--35)e-04	&	(5.7--51)e-05	&	(5.7--51)e-05	\\
Characteristic velocity (km s$^{-1}$)							&	0.8		         &	1.5		         &	2.1		         &	4.0		\\
Size (AU)												&	1000		         &	1200		         &	800		         &	920		\\
Dynamical Time (yr)										&	6200		         &	3800		         &	1800		         &	1100		\\
Momentum ($M_{\odot}$ km s$^{-1}$)						&	(3.0--27)e-04	&	(5.9--53)e-04	&	(1.2--11)e-04	&	(2.3--21)e-04	\\
Kinetic energy ($M_{\odot}$ km$^{2}$ s$^{-2}$)					&	(1.3--12)e-04	&	(5.3--48)e-04	&	(1.2--11)e-04	&	(4.8--43)e-04	\\
Outflow force ($M_{\odot}$ km s$^{-1}$ yr$^{-1}$)				&	(4.9--44)e-08    &	(1.6--14)e-07	&	(6.6--61)e-08	&	(2.1--19)e-07	\\
Mechanical luminosity ($L_{\odot}$)							&	(3.0--28)e-06	&	(1.9--17)e-05	&	(1.1--10)e-05	&	(6.9--62)e-05	\\
Mass Loss rate ($M_{\odot}$ yr$^{-1}$)						&	(6.3--57)e-08	&	(1.0--9.2)e-07	&	(3.1--28)e-08	&	(5.2--47)e-08	\\
\enddata
\tablenotetext{a}{The inclination angle is not corrected}
\tablenotetext{b}{The inclination angle is corrected. The outflow inclination angle was assumed as 30$^{\circ}$.}
\tablenotetext{c}{The outflow mass is estimated using the blueshifted emission assuming with the optically thin gas and the LTE condition.}
\tablenotetext{d}{The lower limit values are derived with optically thin conditions, while the upper limit values are derived with the opacity of $\sim$9. 
After Section 4.2, the upper limit values are used for the outflow discussions.}
\end{deluxetable}

\end{document}